\documentclass[aps,prd,tightenlines,preprint,nofootinbib,a4paper]{revtex4}
\usepackage{epsfig}
\usepackage{amsmath}
\usepackage{appendix}
\usepackage{slashbox}

\begin{document}

\title{Eigenvector-eigenvalue identities and an application to flavor physics}

\author{S. H. Chiu\footnote{schiu@mail.cgu.edu.tw}}
\affiliation{Physics Group, CGE, Chang Gung University, 
Taoyuan 33302, Taiwan}

\author{T. K. Kuo\footnote{tkkuo@purdue.edu}}
\affiliation{Department of Physics, Purdue University, West Lafayette, IN 47907, USA}

\begin{abstract}

The eigenvector-eigenvalue identities are expanded to include general mixing parameters.
Some simple relations are obtained and they reveal an intricate texture of connections
between the eigenvalues and the mixing parameters.
Permutation symmetry ($S_{3}\times S_{3}$) plays an indispensable role in our analysis.
It is the guiding principle for the understanding of our results -- all of them are tensor
equations under permutation.

\end{abstract}


\maketitle

\pagenumbering{arabic}



\section{Introduction}

The diagonalization of hermitian matrices follows a well-established procedure.
One solves for the eigenvalues and eigenvectors. The latter are then collected as a unitary matrix,
which becomes the mixing matrix for the diagonalization.
It is thus very interesting, and surprising, that an alternative method, building on earlier researches,
was discovered \cite{Denton1,Denton2} recently.  
It was found that the mixing matrix elements (squared, $|V_{\alpha i}|^{2}$) can be 
elegantly expressed in terms of the eigenvalues of the hermitian matrix and those of its minors (also hermitian)
along the diagonal. These results were proved for $n \times n$ matrices.

In this paper we will examine in detail the specific case of $3 \times 3$ hermitian matrices,
which, as fermion mass matrices, are fundamental elements in flavor physics.
There are several considerations which suggest this investigation.
Mathematically, the intricacy of the problem starts to show up for $n\geq 3$, 
while the cases $n=2$ may be regarded as a degenerate case of the full-fledged problem.
Physically, in actual applications our knowledge about $|V_{\alpha i}|^{2}$ and the matrix itself
is often piecemeal and uneven. And one frequently employs mixing variables instead of the
matrix elements. It is thus useful to find relations directly between eigenvalues and mixing variables.
Given the wealth of existing informations about flavor physics, it seems interesting to
investigate the detailed roles played by the various concrete physical parameters.
As we will find out, in a diagonalization problem, there are three classes of variables:
1) Those related to flavor, with labels $(\alpha,\beta,\gamma)$; 2) Those with indices $(i.j.k)$,
which belong to mass-eigenstates; 3) Mixing parameters which carry both sets of indices. 
The eigenvector-eigenvalue identities manage to express the mixing parameters directly
in terms of the first two, where the mixing parameters include not just $|V_{\alpha i}|^{2}$,
but also other interesting combinations constructed out of the $V_{\alpha i}$'s. 
Some of our results are simple and suggestive, and may lead 
to new insights about flavor physics.
\section{Notations and mathematical preliminaries}

Although the diagonalization of hermitian matrices applies to the general problem of the mixing of quantum
states, for definiteness, we will use the concrete example of neutrino mixing in this paper.
Thus, we concentrate on the (effective) neutrino mass matrix, $[M]_{\alpha \beta}$ 
($(\alpha,\beta)=(e,\mu,\tau)$), which is diagonalized by the mixing matrix $V_{\alpha i}$,
$V^{\dag}MV=M_{D}$, where the diagonal matrix has eigenvalues $\lambda_{i}$ $(i=1,2,3)$.
With rephasing invariance, the physical observables are constructed out of the combinations of $V_{\alpha i}$.
It was found \cite{KuoLee}  that 
a set of basic, rephasing invariant combinations (RIC) are given by
\begin{equation}\label{Gamma}
\Gamma^{ABC}_{IJK}=E_{ABC}E_{IJK}V_{AI}B_{BJ}V_{CK}    \textrm{     (no sum)},
\end{equation}
where $E_{IJK}(E_{ABC})$ is the symmetric Levi-Civita symbol \cite{KC1} 
which is symmetric under index exchanges, and
\begin{equation}
E_{IJK}=\left\{\begin{array}{ll}
1,  & I\neq J \neq K \\
0, & \mbox{otherwise}
\end{array}
\right.
\end{equation}
Also, without loss of generality, we demand that $\det V=+1$, and thus
\begin{equation}\label{Vai}
V^{*}_{\alpha i}=\frac{1}{2}e_{\alpha \beta \gamma}e_{ijk}V_{\beta j}V_{\gamma k}.
\end{equation}
These $\Gamma's$ have a common imaginary part $(-\mathcal{J})$, 
with $\mathcal{J}$ the Jarlskog invariant \cite{J}, 
while their real parts are defined by 
$(x_{1},x_{2},x_{3};y_{1},y_{2},y_{3})=(\Gamma^{e,\mu,\tau}_{123},\Gamma^{e,\mu,\tau}_{231},
\Gamma^{e,\mu,\tau}_{312},\Gamma^{e,\mu,\tau}_{132},\Gamma^{e,\mu,\tau}_{213},\Gamma^{e,\mu,\tau}_{321}).$
Thus, e.g., $\Gamma^{e\mu \tau}_{123}=x_{1}-i\mathcal{J}$, etc. 
The notation \cite{KuoLee} used here for $(x_{i},y_{j})$ is unfortunate. 
The index $``i"$ does not indicate their transformation properties
under $S_{3}\times S_{3}$. They can, however, be inferred from 
Eqs.(\ref{x1bar}-\ref{y3bar}).

The absolute squares of 
$V_{\alpha i}$ are defined as
\begin{equation}
W_{\alpha i}=|V_{\alpha i}|^{2},
\end{equation}
which are given by the differences of two $\Gamma's$, so that
\begin{equation}\label{W}
[W]=
\left(\begin{array}{ccc}
   x_{1}-y_{1}& x_{2}-y_{2} & x_{3}-y_{3} \\
   x_{3}-y_{2} & x_{1}-y_{3} & x_{2}-y_{1} \\
   x_{2}-y_{3}& x_{3}-y_{1} & x_{1}-y_{2} \\
    \end{array}
    \right). 
\end{equation}
It is also found that $W's$ cofactor matrix, $w^{T}W=Ww^{T}=\det W$, is given by
\begin{equation}\label{w}
[w]=
\left(\begin{array}{ccc}
   x_{1}+y_{1}& x_{2}+y_{2} & x_{3}+y_{3} \\
   x_{3}+y_{2} & x_{1}+y_{3} & x_{2}+y_{1} \\
   x_{2}+y_{3}& x_{3}+y_{1} & x_{1}+y_{2} \\
    \end{array}
    \right). 
\end{equation}
The unitarity condition on $W$ is satisfied by the consistency condition:
\begin{equation}
\Sigma x_{i}-\Sigma y_{j}=\det V=+1,
\end{equation}
while another consistency condition is
\begin{equation}
\Sigma' x_{i}x_{j}=\Sigma' y_{i}y_{j}, (\Sigma'=\Sigma_{i<j}),
\end{equation}
so that there are four independent parameters in the set $(x_{i},y_{j})$.

Other interesting physical variables can be constructed out of $(x_{i},y_{j})$. 
E.g., to evaluate $\det W$, one can replace any row or column in the determinant by
$(1,1,1)$ from unitarity, thus
\begin{eqnarray}\label{DW}
\mathcal{D}=\det W&=&\Sigma_{\alpha}w_{\alpha i}=\Sigma_{i}w_{\alpha i}  \nonumber \\
 &=&\Sigma x_{i}+\Sigma y_{i}.
 \end{eqnarray}
Some other interesting combinations are
\begin{equation}
\Sigma x_{i}=\frac{1}{2}(\mathcal{D}+1),
\end{equation}
\begin{equation}
\Sigma y_{i}=\frac{1}{2}(\mathcal{D}-1),
\end{equation}
and, in particular,
\begin{equation}\label{J2}
\mathcal{J}^{2}=\Pi x_{i}-\Pi y_{i}.
\end{equation}

We now turn to summarizing the behavior of these variables under permutation,
either in the flavor basis $(\alpha,\beta,\gamma)$ or in the mass-eigenvalue
basis $(i,j,k)$. For $V_{\alpha i}$, under exchange operators $(\alpha \leftrightarrow \beta)$
or $(i\leftrightarrow j)$, there is an undetermined phase. To keep $\det V=+1$,
we choose the phase factor (-1). There remain possible additional phases which we can
remove by rephasing. Thus, the transformation laws are
\begin{equation}
V_{\alpha i} \longrightarrow -V_{\beta i}, (\alpha \leftrightarrow \beta)
\end{equation}

\begin{equation}
V_{\alpha i} \longrightarrow -V_{\alpha j}, (i \leftrightarrow j)
\end{equation}
which remain intact as long as we only use $V_{\alpha i}$ in rephasing invariant combinations (RIC).
From these the transformation laws of physical variables (RIC) can be read off directly from their labels.

It may be further noticed that the basic units $\Gamma^{\alpha \beta \gamma}_{ijk}$ and 
$W_{\alpha i}$ (with $V^{*}_{\alpha i}$ given in Eq.(\ref{Vai}))   
contain each flavor and eigenvalue index once, and only once.
This means that physical variables composed out of them have the same property.
Or, under permutation of the indices, the physical variables are tensors.
A direct consequence is that all the relations that we obtain are tensor equations under permutation.

\section{Mixing parameters}

In Ref.\cite{Denton1}, elegant identities were established which connect the mixing element $W_{\alpha i}$
directly with the eigenvalues of an $n \times n$ hermitian matrix and those of its
$(n-1) \times (n-1)$ submatrices along the diagonal. We will now examine its detailed
implementation to $3\times 3$ matrices.
To do this we adopt the concrete notation of neutrino mass matrix,
$[M]_{\alpha \beta}$, $(\alpha,\beta)=(e,\mu,\tau)$, with the diagonal submatrices $[M_{\alpha}]$,
which are obtained from $[M]_{\alpha \beta}$ by deleting the $\alpha$-th row and column.
The eigenvalues of $[M]_{\alpha \beta}$ and $[M_{\alpha}]$
are denoted as $\lambda_{i}$ and $(\xi_{\alpha},\eta_{\alpha})$, respectively, with
$i=(1,2,3)$ and $\alpha=(e,\mu,\tau)$. The diagonalizing mixing matrix is $V_{\alpha i}$, with
$W_{\alpha i}=|V_{\alpha i}|^{2}$. It is convenient to define the traces and determinants of the 
submatrices as
\begin{equation}\label{t-alpha}
t_{\alpha}=\xi_{\alpha}+\eta_{\alpha}
\end{equation}
\begin{equation}\label{d-alpha}
d_{\alpha}=\xi_{\alpha}\eta_{\alpha}  \textrm{     (no sum)}.
\end{equation}
They are not independent and satisfy two constraint conditions:
\begin{equation}\label{ta}
\Sigma_{\alpha}t_{\alpha}=2\Sigma_{i}\lambda_{i},
\end{equation}
\begin{equation}\label{da}
\Sigma_{\alpha}d_{\alpha}=\Sigma'\lambda_{i}\lambda_{j}=\Sigma_{i<j}\lambda_{i}\lambda_{j},
\end{equation}
where the second relation can be proved by an expansion of $\det(\lambda I -M)=\Pi (\lambda-\lambda_{i})$.
There are thus four independent parameters in the set $(t_{\alpha},d_{\alpha})$,
which agrees with the number of independent variables from the set $W_{\alpha i}$.
In other words, the seven eigenvalues contained in $M$ and $M_{\alpha}$ ($\det M$, $d_{\alpha}$, $t_{\alpha}$)
can be divided into two groups.
One set ($\det M=\Pi \lambda_{l}$, $\Sigma d_{\alpha}$, $\Sigma t_{\alpha}$) determines $\lambda_{i}$,
while the rest morph (together with $\lambda_{i}$) into mixing parameters which bridge the flavor
space and mass eigenvalue space.

The eigenvector-eigenvalue identities are given by \cite{Denton2} 
\begin{equation}\label{WDenton}
W_{\alpha i}=(\lambda_{i}-\xi_{\alpha})(\lambda_{i}-\eta_{\alpha})/(\lambda_{i}-\lambda_{j})(\lambda_{i}-\lambda_{k}).
\end{equation}
To elucidate their transformation properties under permutation 
in either the flavor or mass eigenvalue basis, it is convenient to introduce the
traceless and anti-symmetric tensors  $(\widetilde{\textbf{3}})$:
\begin{equation}
\widetilde{\lambda}_{i}=\frac{1}{2}e_{ijk}(\lambda_{j}-\lambda_{k}),
\end{equation}
and
\begin{equation}
\widetilde{\lambda}_{i}\lambda_{i}=((\lambda_{2}-\lambda_{3})\lambda_{1},(\lambda_{3}-\lambda_{1})\lambda_{2},
(\lambda_{1}-\lambda_{2})\lambda_{3}),
\end{equation}
and a pseudoscalar $(\widetilde{\textbf{1}})$:
\begin{equation}
\Pi  \widetilde{\lambda}_{l}=(\lambda_{1}-\lambda_{2})(\lambda_{2}-\lambda_{3})(\lambda_{3}-\lambda_{1}).
\end{equation}
The trace of another tensor $\widetilde{\lambda}_{i}\lambda_{i}^{2}$  $(\widetilde{\textbf{3}})$ is given by
\begin{equation}
\Sigma \widetilde{\lambda}_{i}\lambda_{i}^{2}=-\Pi \widetilde{\lambda}_{l}.
\end{equation}
Thus, we have
\begin{equation}\label{Walphai}
W_{\alpha i}=-(\Pi \widetilde{\lambda}_{l})^{-1}[\widetilde{\lambda}_{i}(\lambda_{i}^{2}-\lambda_{i}t_{\alpha}+d_{\alpha})],
\end{equation}
which exhibits its transformation properties explicitly.

Let us pause and analyze some of the implications of these results.
The transformation from the flavor basis ($F$-basis) to the mass-eigenvalue basis ($M$-basis)
is mediated by $W_{\alpha i}$, which are now given explicitly in terms of the eigenvalues.
As noted in Ref.\cite{Denton1}, Eqs.(\ref{WDenton}) and (\ref{Walphai}) are invariant under the 
following transformations:

1) Translation:  $M \rightarrow M+\delta[I]$,
\begin{equation}
(\lambda_{i};\xi_{\alpha},\eta_{\alpha};t_{\alpha},d_{\alpha};W_{\alpha i}) \rightarrow(\lambda_{i}+\delta;\xi_{\alpha}+\delta,
\eta_{\alpha}+\delta; t_{\alpha}+2\delta, d_{\alpha}+\delta t_{\alpha}+\delta^{2};W_{\alpha i}). \nonumber
\end{equation} 

2) Dilatation: $M \rightarrow rM$,
\begin{equation}
(\lambda_{i};\xi_{\alpha},\eta_{\alpha};t_{\alpha},d_{\alpha};W_{\alpha i}) \rightarrow(r\lambda_{i};r\xi_{\alpha},
r\eta_{\alpha}; rt_{\alpha}, r^{2}d_{\alpha};W_{\alpha i}). \nonumber
\end{equation} 
Invariance under dilatation means that a dimension analysis can be effective in constraining
the forms taken by the (dimensionless) mixing parameters, as can be seen in all of the results that follow.

3) Permutation:
\begin{equation}
[S_{3}]_{F}: (M_{\alpha \beta} \leftrightarrow M_{\beta \alpha}; t_{\alpha}\leftrightarrow t_{\beta};
d_{\alpha} \leftrightarrow d_{\beta}, W_{\alpha i} \leftrightarrow W_{\beta i}), \nonumber
\end{equation}
\begin{equation}
[S_{3}]_{M}:(\lambda_{i}\leftrightarrow \lambda_{j};W_{\alpha i}\leftrightarrow W_{\alpha j}). \nonumber
\end{equation}
Physically, as noted in Ref.\cite{KC1,KC2}, the symmetry $[S_{3}]_{F} \times [S_{3}]_{M}$
stems from the freedom to reorder the states in a diagonalization process.
Physical variables $(\lambda_{i},W_{\alpha i}, etc.)$ transform as tensors, the Lagrangian of the
standard model is invariant, and evolution equations of the physical variables are tensor equations
under $[S_{3}]_{F} \times [S_{3}]_{M}$. The eigenvector-eigenvalue identities corroborate 
this analysis and, we believe, help to put permutation symmetry on firm grounds.

Turning now to other mixing variables, we start with the cofactor matrix $[w]$, which has elements
\begin{equation}
w_{\alpha i}=\frac{1}{2!}e_{\alpha \beta \gamma}e_{ijk}W_{\beta j}W_{\gamma k}.
\end{equation}
Using Eq.(\ref{Walphai}), we find 
\begin{equation}
(\Pi \widetilde{\lambda}_{l})w_{\alpha i}=-(\lambda_{j}+\lambda_{k})\widetilde{d}_{\alpha}+
\lambda_{j}\lambda_{k}\widetilde{t}_{\alpha}+(d_{\beta} t_{\gamma}-d_{\gamma} t_{\beta}),
\end{equation}
where $(i,j,k)$ and $(\alpha,\beta,\gamma)$ are cyclic permutations of the bases, and we have constructed the
anti-symmetric combinations
\begin{equation}
\widetilde{t}_{\alpha}=\frac{1}{2!}e_{\alpha \beta \gamma}(t_{\beta}-t_{\gamma}),
\end{equation}
\begin{equation}
\widetilde{d}_{\alpha}=\frac{1}{2!}e_{\alpha \beta \gamma}(d_{\beta}-d_{\gamma}).
\end{equation}
Using Eqs.(\ref{t-alpha}) and (\ref{d-alpha}), plus the definition
\begin{equation}
\widetilde{\xi}_{\alpha}=\frac{1}{2}e_{\alpha \beta \gamma}(\xi_{\beta}-\xi_{\gamma}),
\end{equation}
\begin{equation}
\widetilde{\eta}_{\alpha}=\frac{1}{2}e_{\alpha \beta \gamma}(\eta_{\beta}-\eta_{\gamma}),
\end{equation}
we can recast $w_{\alpha i}$ in the form
\begin{equation}
(\Pi \widetilde{\lambda}_{l})w_{\alpha i}=\widetilde{\xi}_{\alpha}(\lambda_{j}-\eta_{\beta})(\lambda_{k}-\eta_{\gamma})
+\widetilde{\eta}_{\alpha}(\lambda_{j}-\xi_{\gamma})(\lambda_{k}-\xi_{\beta}),
\end{equation}
or, equivalently,
\begin{equation}
(\Pi \widetilde{\lambda}_{l})w_{\alpha i}=\widetilde{\xi}_{\alpha}(\lambda_{j}-\eta_{\gamma})(\lambda_{k}-\eta_{\beta})
+\widetilde{\eta}_{\alpha}(\lambda_{j}-\xi_{\beta})(\lambda_{k}-\xi_{\gamma}),
\end{equation}
mimicking Eq. (\ref{WDenton}). Also, note that 
\begin{equation}
\widetilde{\xi}_{\alpha}\eta_{\beta}\eta_{\gamma}+\widetilde{\eta}_{\alpha}\xi_{\beta}\xi_{\gamma}=
d_{\beta}t_{\gamma}-d_{\gamma}t_{\beta},
\end{equation}
and 
\begin{equation}
\Sigma (d_{\beta}t_{\gamma}-d_{\gamma}t_{\beta})=\Sigma d_{\alpha}\widetilde{t}_{\alpha}=
-\Sigma \widetilde{d}_{\alpha}t_{\alpha}. 
\end{equation}
Thus
\begin{eqnarray}\label{D3}
\mathcal{D}&=&\det W=\Sigma w_{\alpha i} \nonumber \\
&=&(\Pi \widetilde{\lambda}_{l})^{-1} \Sigma d_{\alpha}\widetilde{t}_{\alpha}=
-(\Pi \widetilde{\lambda}_{l})^{-1} \Sigma t_{\alpha}\widetilde{d}_{\alpha} \nonumber \\
&=&\frac{1}{2!}(\Pi \widetilde{\lambda}_{l})^{-1}[\Sigma E_{\alpha \beta \gamma}(\widetilde{\xi}_{\alpha}\eta_{\beta}\eta_{\gamma}
+\widetilde{\eta}_{\alpha}\xi_{\beta}\xi_{\gamma})],
\end{eqnarray}
which is a remarkably simple relation between $\det W$ and the eigenvalues. Note also that it is the ratio
of two pseudoscalars, one in flavor space and the other in mass eigenvalue space
($\mathcal{D} \sim (\widetilde{\textbf{1}})_{F} \times (\widetilde{\textbf{1}})_{M}$), in agreement with the
transformation property of $\det W$ under $(S_{3})_{F} \times (S_{3})_{M}$.

It is instructive to compare this result with that of the $2 \times 2$ matrices. There, the eigenvalue solution 
is well-known. Without loss of generality, we may take the matrix to be traceless and real,
\begin{equation}
\left(\begin{array}{cc}
   M_{\alpha \alpha} & M_{\alpha \beta}  \\ \nonumber
   M_{\alpha \beta} & -M_{\alpha \alpha}   \nonumber
      \end{array}
  \right),
\end{equation}
whose eigenvalues are given by 
\begin{equation}
\lambda_{1,2}=\pm \sqrt{M^{2}_{\alpha \alpha}+M^{2}_{\alpha \beta}},
\end{equation}
and the mixing parameters are 
\begin{equation}
W_{e1}-W_{\mu 1}=M_{\alpha \alpha}/\sqrt{M^{2}_{\alpha \alpha}+M^{2}_{\alpha \beta}},
\end{equation}
which agrees with the eigenvector-eigenvalue identity \cite{Denton2},  
\begin{equation}
W_{e1}-W_{\mu 1}=(-\xi_{\alpha}+\xi_{\beta})/(\lambda_{1}-\lambda_{2}).
\end{equation}
In our notation, 
\begin{equation}
\xi_{\alpha}=t_{\alpha}=-M_{\alpha \alpha},
\end{equation}
\begin{equation}
\det W=W_{e1}-W_{\mu 1}=\mathcal{D}_{(2)},
\end{equation}
we have
\begin{equation}\label{D2}
\mathcal{D}_{(2)}=-(t_{\alpha}-t_{\beta})/(\lambda_{1}-\lambda_{2}).
\end{equation}
Since the parameter $\mathcal{D}_{(2)}$ completely determines the mixing pattern, 
Eq.(\ref{D2}) gives an intuitive and elegant formula connecting the three sets of variables
(the dimensionless mixing parameter, the eigenvalues in the flavor, and
the mass eigen-basis) involved in the diagonalization.
Note also that, written in this form, Eq.(\ref{D2}) is valid for an arbitrary $2 \times 2$ hermitian matrix.

Eq.(\ref{D2}) can also be deduced from Eq.(\ref{D3}). 
To do this we consider a degenerate $3\times 3$ matrix for which the elements in the third row and column
vanish, $M_{\gamma \rho}=M_{\rho \gamma}=0$, $\rho=(\alpha, \beta,\gamma)$.
In this case we have $\lambda_{3}=0$, $d_{\alpha}=d_{\beta}=0$, $d_{\gamma}=\lambda_{1}\lambda_{2}$ and
$\widetilde{t}_{\gamma}=t_{\alpha}-t_{\beta}=-2M_{\alpha \alpha}$. 
Then, $\Sigma d_{\alpha}\widetilde{t}_{\alpha}=\lambda_{1}\lambda_{2}(t_{\alpha}-t_{\beta})$,
$\Pi \widetilde{\lambda}_{l}=-\lambda_{1}\lambda_{2}(\lambda_{1}-\lambda_{2})$,
and $\mathcal{D}_{(3)}$ reduces to $\mathcal{D}_{(2)}$. The unmistakable lineage between $\mathcal{D}_{(2)}$ and
$\mathcal{D}_{(3)}$ is thus established.

We now turn to other parameters which are composed out of $W_{\alpha i}$ and $w_{\alpha i}$.
To do this it is convenient to separate the traceless parts (corresponding to extracting
the singlet out of a triplet under $S_{3}$) of these variables, we define 
\begin{equation}
\overline{W}_{\alpha i}=W_{\alpha i}-\frac{1}{3},
\end{equation}
\begin{equation}
\overline{w}_{\alpha i}=w_{\alpha i}-\frac{\mathcal{D}}{3}.
\end{equation}
They are related to $W_{\alpha i}$ and $w_{\alpha i}$ as follows:
\begin{eqnarray}\label{Ww}
3\overline{W}_{\alpha i}&=&w_{\beta j}-w_{\beta k}-w_{\gamma j}+w_{\gamma k} \nonumber \\
&=&\frac{1}{2!}e_{\alpha \beta \gamma}e_{ijk}(\overline{w}_{\beta j}+\overline{w}_{\gamma k}),
\end{eqnarray}
\begin{eqnarray}\label{wW}
3\overline{w}_{\alpha i}&=&W_{\beta j}-W_{\beta k}-W_{\gamma j}+W_{\gamma k} \nonumber \\
&=&\frac{1}{2!}e_{\alpha \beta \gamma}e_{ijk}(\overline{W}_{\beta j}+\overline{W}_{\gamma k}).
\end{eqnarray}
which can be verified using Eqs.(\ref{W}) and (\ref{w}).
Note that, by using traceless variables, these relations are now linear, while those between
$W_{\alpha i}$ and $w_{\alpha i}$ are quadratic,
\begin{equation}
w_{\alpha i}=\frac{1}{2!}e_{\alpha \beta \gamma}e_{ijk}W_{\beta j}W_{\gamma k}, 
\end{equation}
\begin{equation}
\mathcal{D}W_{\alpha i}=\frac{1}{2!}e_{\alpha \beta \gamma}e_{ijk}w_{\beta j}w_{\gamma k}.  
\end{equation}
Eqs.(\ref{Ww}-\ref{wW}) exhibit manifest reciprocity between $\overline{W}_{\alpha i}$ and
$\overline{w}_{\alpha i}$, which however, is camouflaged by the relation between $W_{\alpha i}$
and $w_{\alpha i}$. These equations are also useful in converting $W$ into $w$ and vice versa.

We may further define
\begin{equation}
m_{\alpha i}=(-\Pi \widetilde{\lambda}_{l})W_{\alpha i}-\widetilde{\lambda}_{i} \lambda_{i}^{2}
=\widetilde{\lambda}_{i}(d_{\alpha}-\lambda_{i}t_{\alpha})
\end{equation}
\begin{eqnarray}\label{nm}
n_{\alpha i}&=&m_{\beta i}-m_{\gamma i}=
\widetilde{\lambda}_{i}(\widetilde{d}_{\alpha}-\lambda_{i}\widetilde{t}_{\alpha}) \nonumber \\
&=& -(\Pi \widetilde{\lambda}_{l})(W_{\beta i}-W_{\gamma i}).
\end{eqnarray}
The traces of $m_{\alpha i}$ are given by
\begin{equation}
\Sigma_{i} m_{\alpha i}=0,
\end{equation}
\begin{eqnarray}
\Sigma_{\alpha}m_{\alpha i}&=&\widetilde{\lambda}_{i}(\Sigma d_{\alpha}-\lambda_{i}\Sigma t_{\alpha}) \nonumber \\
&=&-(\Pi \widetilde{\lambda}_{l})-3\widetilde{\lambda}_{i}\lambda^{2}_{i},
\end{eqnarray}
where Eqs. (\ref{ta}) and (\ref{da}) are used. They can be used to verify the unitarity conditions
\begin{equation}
\Sigma_{i}W_{\alpha i}=\Sigma_{\alpha}W_{\alpha i}=1.
\end{equation}
For $n_{\alpha i}$, it is clear that they are traceless with respect to both sets of indices,
\begin{equation}
\Sigma_{\alpha}n_{\alpha i}=\Sigma_{i}n_{\alpha i}.
\end{equation}
Also,
\begin{equation}\label{nab}
n_{\alpha i}-n_{\beta i}=3(\Pi \widetilde{\lambda}_{l})\overline{W}_{\gamma i},
\end{equation}
\begin{equation}
n_{\alpha i}-n_{\alpha j}=-3(\Pi \widetilde{\lambda}_{l})\overline{w}_{\alpha k}.
\end{equation}
Note that $n_{\alpha i}$ is invariant under translation
$(\lambda_{i},\xi_{\alpha},\eta_{\alpha}) \rightarrow (\lambda_{i}+\delta,\xi_{\alpha}+\delta,\eta_{\alpha}+\delta)$,
just like $W_{\alpha i}$, $w_{\alpha i}$, etc.



\begin{table}
\centering
\begin{tabular}
{|c|c||c|c|c|} \hline
 & &$(i,j)=(2,1)$ &$(i,j)=(3,2)$ &$(i,j)=(1,3)$ \\
  \cline{2-5} \cline{2-5}
   & $\overline{x}_{1}$ & $n_{\alpha 2}-n_{\beta 1}$ & $n_{\beta 3}-n_{\gamma 2}$ & $n_{\gamma 1}-n_{\alpha 3}$  \\ 
 \cline{2-5}
 & $\overline{x}_{2}$ & $n_{\gamma 2}-n_{\alpha 1}$ & $n_{\alpha 3}-n_{\beta 2}$ & $n_{\beta 1}-n_{\gamma 3}$   \\ 
 \cline{2-5}
  &$\overline{x}_{3}$ & $n_{\beta 2}-n_{\gamma 1}$ & $n_{\gamma 3}-n_{\alpha 2}$ & $n_{\alpha 1}-n_{\beta 3}$ \\ 
 \cline{2-5}
 $(-3\Pi\widetilde{\lambda}_{l})\times$& $\overline{y}_{1}$ & $n_{\alpha 2}-n_{\gamma 1}$ & $n_{\gamma 3}-n_{\beta 2}$ & $n_{\beta 1}-n_{\alpha 3}$ \\ 
 \cline{2-5}
 & $\overline{y}_{2}$ & $n_{\beta 2}-n_{\alpha 1}$ & $n_{\alpha 3}-n_{\gamma 2}$ & $n_{\gamma 1}-n_{\beta 3}$   \\ 
 \cline{2-5}
 &$\overline{y}_{3}$ & $n_{\gamma 2}-n_{\beta 1}$ & $n_{\beta 3}-n_{\alpha 2}$ & $n_{\alpha 1}-n_{\gamma 3}$  \\ 
 \cline{2-5}  \hline
\end{tabular}
\caption{Three equivalent forms for $(-3\Pi\widetilde{\lambda}_{l})\times$ $(\overline{x}_{i};\overline{y}_{i})$, 
arranged in the order of $n_{\alpha i}-n_{\beta j}$ with $(i,j)=[(2,1);(3,2);(1,3)]$.}
\end{table}

Turning to the $(x,y)$ variables, we again define the traceless variables
\begin{equation}
\overline{x}_{i}=x_{i}-\frac{\mathcal{D}+1}{6},
\end{equation}
\begin{equation}
\overline{y}_{i}=y_{i}-\frac{\mathcal{D}-1}{6},
\end{equation}
we find, e.g.,
\begin{equation}
W_{e1}+W_{\mu 2}+W_{\tau 3}=3\overline{x}_{1}+1,
\end{equation}
or collectively,
\begin{equation}\label{x1bar}
3\overline{x}_{1}=\overline{W}_{e1}+\overline{W}_{\mu 2}+\overline{W}_{\tau 3},
\end{equation}

\begin{equation}
3\overline{x}_{2}=\overline{W}_{e2}+\overline{W}_{\mu 3}+\overline{W}_{\tau 1},
\end{equation}

\begin{equation}
3\overline{x}_{3}=\overline{W}_{e3}+\overline{W}_{\mu 1}+\overline{W}_{\tau 2},
\end{equation}

\begin{equation}
3\overline{y}_{1}=\overline{W}_{e1}+\overline{W}_{\mu 3}+\overline{W}_{\tau 2},
\end{equation}

\begin{equation}
3\overline{y}_{2}=\overline{W}_{e2}+\overline{W}_{\mu 1}+\overline{W}_{\tau 3},
\end{equation}

\begin{equation}\label{y3bar}
3\overline{y}_{3}=\overline{W}_{e3}+\overline{W}_{\mu 2}+\overline{W}_{\tau 1}.
\end{equation}
We can write these variables in terms of $n_{\alpha i}$, in three equivalent ways, 
as shown in Table I.

Combined with the expression for $\mathcal{D}$, Eq.(\ref{D3}), the basic variables $(x_{i},y_{j})$ can be written
in terms of the eigenvalues, and so can the mixing parameters constructed out of $(x_{i},y_{j})$.
We will now present a concrete example as follows. 

It was pointed out that permutation symmetry suggests the use 
of singlets as mixing parameters \cite{KC2}.
In addition to $\mathcal{D}$ and $\mathcal{J}^{2}$, as defined in Eqs.(\ref{DW}) and (\ref{J2}), respectively,
one may propose another two parameters,
\begin{equation}
\mathcal{Q}^{2}=\Sigma x_{i}x_{j}+\Sigma y_{i}y_{j},
\end{equation}
\begin{equation}
\mathcal{K}=\Pi x_{i}+\Pi y_{i},
\end{equation}
to form a set of four parameters, which may serve to characterize flavor mixing.
The set $(\mathcal{D},\mathcal{Q}^{2},\mathcal{K},\mathcal{J}^{2})$ is unique in that it does not
contain superfluous variables.
This situation is similar to that in relativistic theories, where it is preferrable to use 
relativistic invariants, rather than frame-dependent variables, as physical parameters.

Before writing them in terms of eigenvalues,
let's first verify the constraint $\Sigma'(x_{i}x_{j}-y_{i}y_{j})=0$. We note that
\begin{equation}
\Sigma'(x_{i}x_{j}-y_{i}y_{j})=\Sigma'(\overline{x}_{i}\overline{x}_{j}-\overline{y}_{i}\overline{y}_{j})+\frac{1}{3}\mathcal{D},
\end{equation}
and 
\begin{equation}
2\Sigma'(\overline{x}_{i}\overline{x}_{j}-\overline{y}_{i}\overline{y}_{j})=-\Sigma (\overline{x}_{i}^{2}-\overline{y}_{i}^{2}).
\end{equation}
Using Eq. (\ref{nab}) and the expressions for $n_{\alpha i}-n_{\beta j}$ in the $(i,j)=(2,1)$ column of Table I, we find
\begin{equation}
\Sigma(\overline{x}_{i}^{2}-\overline{y}_{i}^{2})
=-\frac{2}{3}(\Pi \widetilde{\lambda}_{l})^{-1}(\Sigma_{\alpha}n_{\alpha 2}W_{\alpha 1}),
\end{equation}
\begin{eqnarray}\label{nW}
\Sigma_{\alpha} n_{\alpha 2}W_{\alpha 1}
&=&-(\Pi \widetilde{\lambda}_{l})^{-1}\widetilde{\lambda}_{1}\widetilde{\lambda}_{2} (\lambda_{1}-\lambda_{2})[\Sigma_{\alpha} d_{\alpha}\widetilde{t}_{\alpha}] \nonumber \\
&=&-(\Pi \widetilde{\lambda}_{l})\mathcal{D}.
\end{eqnarray}
Thus, $\Sigma'(x_{i}x_{j}-y_{i}y_{j})=0$.
Note that, had we used $(i,j)=(3,2)$ or $(1,3)$ to calculate $\mathcal{D}$ in Eq.(\ref{nW}), 
we would have gotten the same result. Thus, we may recast Eq.(\ref{nW}) as
\begin{equation}\label{piD}
(\Pi \widetilde{\lambda}_{l}) \mathcal{D}=\frac{1}{3}\Sigma'_{i<j}\Sigma_{\alpha} (n_{\alpha i}W_{\alpha j}),
\end{equation}
showing explicitly that $\mathcal{D}$ is a singlet under $S_{3} \times S_{3}$.

In addition to the expression for $\mathcal{D}$ as function of eigenvalues, as in Eq.(\ref{D3}), 
the calculation of $Q^{2}=\Sigma'(x_{i}x_{j}+y_{i}y_{j})$ is more involved. Start from
\begin{equation}
\Sigma'(x_{i}x_{j}+y_{i}y_{j})
=\Sigma'(\overline{x}_{i}\overline{x}_{j}+\overline{y}_{i}\overline{y}_{j})+\frac{\mathcal{D}^{2}+1}{6},
\end{equation}
\begin{equation}
\Sigma'(\overline{x}_{i}\overline{x}_{j}+\overline{y}_{i}\overline{y}_{j})
=-\frac{1}{2}\Sigma(\overline{x}_{i}^{2}+\overline{y}_{i}^{2}),
\end{equation}
and calculating $\Sigma(\overline{x}_{i}^{2}+\overline{y}_{i}^{2})$ by using the average of the three columns in Table I, 
we find
\begin{equation}
\frac{9}{2}(\Pi \widetilde{\lambda}_{l})^{2} \cdot \Sigma (\overline{x}_{i}^{2}+\overline{y}_{i}^{2})
=-\frac{1}{3}\Sigma'_{i<j}\Sigma_{\alpha}(n_{\alpha i}n_{\alpha j}).
\end{equation}
Another way to display this result is to use Eq.(\ref{nm}) and write it explicitly in terms of
$(\lambda_{i},d_{\alpha},t_{\alpha})$. The result is
\begin{eqnarray}\label{Q2}
\mathcal{Q}^{2}&=&\frac{1}{27}(\Pi \widetilde{\lambda}_{l})^{-2}[\Sigma'_{i<j}\Sigma_{\alpha}(n_{\alpha i}n_{\alpha j})]
+\frac{\mathcal{D}^{2}+1}{6} \nonumber \\
&=&\frac{1}{27}(\Pi \widetilde{\lambda}_{l})^{-2}\{[-\frac{1}{4}(\Sigma t_{\alpha})^{2}+ \Sigma d_{\alpha}][\Sigma (\widetilde{d}_{\alpha})^{2}] \nonumber \\
&+&[-\frac{1}{2}(\Sigma t_{\alpha})(\Sigma d_{\alpha})+9(\Pi \widetilde{\lambda}_{l})]
[\Sigma (\widetilde{d}_{\alpha}\widetilde{t}_{\alpha})] \nonumber \\
&+&[(\Sigma d_{\alpha})^{2}-\frac{1}{2}(\Sigma t_{\alpha})(\Pi \widetilde{\lambda}_{l})] [\Sigma (\widetilde{t}_{\alpha})^{2}]\}
+\frac{\mathcal{D}^{2}+1}{6}.
\end{eqnarray}


\begin{table}
\centering
\begin{tabular}
{c||c|c} 
 & $(x_{i},y_{j})$& Eigenvalues \\
 \hline \hline
 $\mathcal{D}$& $\Sigma x_{i}+\Sigma y_{i}$&
 $\frac{1}{3}(\Pi \widetilde{\lambda}_{l})^{-1} [\Sigma'_{i<j}\Sigma_{\alpha} (n_{\alpha i}W_{\alpha j})]$ 
 $=(\Pi \widetilde{\lambda}_{l})^{-1}\Sigma d_{\alpha}\widetilde{t}_{\alpha}
 =-(\Pi \widetilde{\lambda}_{l})^{-1}\Sigma t_{\alpha}\widetilde{d}_{\alpha}$\\
 \hline 
 $\mathcal{Q}^{2} $ & $\Sigma (x_{i}x_{j}+y_{i}y_{j})$ 
 &$ \frac{1}{27}(\Pi \widetilde{\lambda}_{l})^{-2}[\Sigma'_{i<j}\Sigma_{\alpha}(n_{\alpha i}n_{\alpha j})]
+\frac{\mathcal{D}^{2}+1}{6}$ \\ 
 \hline
$\mathcal{K}$ & $\Pi x_{i}+\Pi y_{i}$ &$(-\frac{1}{9})(\Pi \widetilde{\lambda}_{l})^{-2}
[\frac{1}{3}\Sigma_{i}\Sigma'_{\alpha \beta \gamma} (n_{\alpha i}n_{\beta i}\overline{w}_{\gamma i})]
+\frac{D}{54}(9\mathcal{Q}^{2}-\mathcal{D}^{2}+3)$   \\ 
\hline
 $\mathcal{J}^{2}$ &$\Pi x_{i}-\Pi y_{i}$ & $(-\frac{1}{9})(\Pi \widetilde{\lambda}_{l})^{-2}[\frac{1}{3}\Sigma_{\alpha}\Sigma'_{ijk} (n_{\alpha i}n_{\alpha j}\overline{W}_{\alpha k})]
+\frac{1}{54}(9\mathcal{Q}^{2}-2\mathcal{D}^{2}-1)$  \\ 
\hline

\end{tabular}
\caption{Expressions for $(\mathcal{D},\mathcal{Q}^{2},\mathcal{K},\mathcal{J}^{2})$ in terms of
$(x_{i},y_{j})$ and the eigenvalues.}
\end{table}

For $\mathcal{K}$ and $\mathcal{J}^{2}$, we will only give the results in terms of $n_{\alpha i}$:
\begin{equation}\label{KK}
\mathcal{K}=(-\frac{1}{9})(\Pi \widetilde{\lambda}_{l})^{-2}
[\frac{1}{3}\Sigma_{i}\Sigma'_{\alpha \beta \gamma} (n_{\alpha i}n_{\beta i}\overline{w}_{\gamma i})]
+\frac{D}{54}(9\mathcal{Q}^{2}-\mathcal{D}^{2}+3),
\end{equation}

\begin{equation}\label{J2J}
\mathcal{J}^{2}=(-\frac{1}{9})(\Pi \widetilde{\lambda}_{l})^{-2}[\frac{1}{3}\Sigma_{\alpha}\Sigma'_{ijk} (n_{\alpha i}n_{\alpha j}\overline{W}_{\alpha k})]
+\frac{1}{54}(9\mathcal{Q}^{2}-2\mathcal{D}^{2}-1),
\end{equation}
where $(\overline{x}_{i},\overline{y}_{i})$ as functions of $n_{\alpha i}$ (and thus of the eigenvalues) 
are given in Table I and Eq.(\ref{nm}), respectively.
Eqs.(\ref{piD}), (\ref{Q2}), (\ref{KK}), and (\ref{J2J}) form a complete set for
$(\mathcal{D},\mathcal{Q}^{2},\mathcal{K},\mathcal{J}^{2})$. We present them in Table II.

In summary, starting from the eigenvector-eigenvalue identities, we can write the mixing parameters in
terms of the eigenvalues in the flavor basis and in mass-eigenvalue basis.
This is facilitated by noticing that $\Sigma x_{i}=\frac{1}{2}(\mathcal{D}+1)$ and $\Sigma y_{i}=\frac{1}{2}(\mathcal{D}-1)$,
with $\mathcal{D}$ given in Eq.(\ref{DW}). The traceless variables $\overline{x}_{i}$ and $\overline{y}_{i}$ are
now linear functions of $W_{\alpha i}$, and can be written as simple functions of the eigenvalues.
Through out these manipulations, permutation symmetry is an indispensable guide, and all relations thus obtained
are tensor equations under permutation.  
Other examples which manifest the permutation symmetry properties of the neutrino
parameters can be found, e.g., in Ref \cite{zhou}.

It should also be mentioned that, given $W_{\alpha i}$, one can construct $\mathcal{J}^{2}$,
but the sign of $\mathcal{J}=\pm\sqrt{\mathcal{J}^{2}}$ is undetermined. 
The sign of $\mathcal{J}$ is fixed by summing up Eq.(\ref{Gamma}), 
(1/3!)$\Sigma \Gamma^{\alpha \beta \gamma}_{ijk}=\mathcal{D}-i(3!)\mathcal{J}$. 
Thus, the variables $W_{\alpha i}$ can only determine CP-violation effects up to a sign.

\section{Conclusion}
In this paper we generalized the eigenvector-eigenvalue identities to include general mixing parameters.
These rephasing invariant variables are tensors under $S_{3} \times S_{3}$.
A well-known set is of course $W_{\alpha i}$, which, however, has a lot of redundancy.
A smaller set is $(x_{i},y_{j})$, with six variables and two consistency conditions.
The most economical set is  $(\mathcal{D},\mathcal{Q}^{2},\mathcal{K},\mathcal{J}^{2})$,
with exactly four parameters. They are all singlets under rephasing  and permutation,
in tune with the symmetry properties of the diagonalization process.
We obtained direct link between these variables and the eigenvalues, summarized in Tables I and II.
Of particular interest may be Eq.(\ref{D3}), which expressed $\mathcal{D}=\det W$ in terms of
the ratio of two eigenvalue expressions. Its use has enabled us to greatly simplify our results.
Together with the other results, they suggest a deep-rooted connection
between the mixing parameters, the flavor eigenvalues, and the mass-eigenvalues.


Our analysis also brings out the prominent role played by permutation symmetry (see also Ref. \cite{Denton1}).  
All of the relations are tensor equations under $[S_{3}]_{F} \times [S_{3}]_{M}$,
without which they would be very hard to decipher. It should be emphasized that the eigenvalue and 
mixing parameters must transform together. This is in contrast to the narrative, that masses are
fixed numbers and not subject to transformations.  However, an inherent property of the mass matrix
diagonalization problem is that it does not predetermine the order of states.
If one makes an exchange of states, it is only natural to also exchange the eigenvalues associated with the states.
The resulting permutation symmetry then reflects the freedom of choice of ordering
in the diagonalization process.

Another indication of the working of permutation symmetry is seen in our maneuvering to get the results.
The basic variables used in this paper transform as $\textbf{3}$'s, which are reducible.
The use of their traceless parts, which are irreducible, has been effective in simplifying our results.

Finally, we add some speculative remarks.
In the standard model, the fermion mass matrices are notorious for being incomprehensible.
At the same time, the observed masses and mixing parameters do seem to have some regularity
(hierarchy). It is hoped that, with the availability of a direct connection between
these entities, some new approach/insight may be uncovered.


\acknowledgments                 
SHC is supported by the Ministry of Science and Technology of Taiwan, 
Grant No.: MOST 109-2112-M-182-001.





\begin{thebibliography}{99}

\bibitem{Denton1}
 P. B. Denton, S. J. Parke, T. Tao, and X. Zhang,
Bull. Am. Math. Soc. {\bf} 59, no.1, 31-58 (2022)
; arXiv:1908.03795v4 [math.RA]  (2019).
The development of the eigenvector-eigenvalue identities has a fascinating history,
with contributions from many authors over a long time. For a summarizing account,
see Sec. 3 of this paper.
  
  \bibitem{Denton2}
P. B. Denton, S. J. Parke, and X. Zhang, 
Phys.\ Rev.\ D {\bf 101}, 093001 (2020)

\bibitem{KuoLee}
T.~K.~Kuo and T.~H.~Lee, 
Phys.\ Rev.\ D {\bf 71}, 093011 (2005)

\bibitem{KC1}
T. K. Kuo and S. H. Chiu, 
Adv. High Energy Phys. {\bf 2020}, 2491897 (2020)

\bibitem{J}
 C. Jarlskog, 
 Phys. Rev. Lett. {\bf 55}, 1039 (1985)


\bibitem{KC2}
T. K. Kuo and S. H. Chiu,
Eur. Phys. J. C {\bf 80} (2020) 3, 203; Eur. Phys. J. C {\bf 80} (2020) 10, 941 (erratum)


\bibitem{zhou}
 Shun Zhou,
 J. Phys. G {\bf 49}, 025004 (2022) 


	
\end{thebibliography}
\end{document}